\documentclass[12pt]{article}
\usepackage{amssymb}

\newcommand{\beq}[1]{\begin{equation}\label{#1}}
\newcommand{\eeq}{\end{equation}}
\newcommand{\bear}[1]{\begin{eqnarray}\label{#1}}
\newcommand{\ear}{\end{eqnarray}}
\newcommand{\nn}{\nonumber}

\catcode`\@=11 \@addtoreset{equation}{section}\catcode`\@=12

 \def\barr{\left(\begin{array}}
 \def\earr{\end{array}\right)}

\newcommand{\N}{ {\mathbb N} }
\newcommand{\R}{ {\mathbb R} }

\newcommand{\diag}{ \mbox{\rm diag} }
\newcommand{\sign}{ \mbox{\rm sign} }
\newcommand{\e}{ \mbox{\rm e} }
\newcommand{\eps}{ \varepsilon }

\newcommand{\p}{\partial}

\newcommand{\tri}{\Delta}

\newcommand{\fnm}{\footnotemark}
\newcommand{\fnt}{\footnotetext}

\begin{document}

 \begin{center}
 \large \bf

  Quantum billiards in multidimensional models with branes

 \end{center}

 \vspace{0.3truecm}

 \begin{center}

 \normalsize\bf

  V. D. Ivashchuk\fnm[1]\fnt[1]{e-mail: ivashchuk@mail.ru}
  and V. N. Melnikov\fnm[2]\fnt[2]{e-mail:  melnikov@phys.msu.ru},

  \vspace{0.5truecm}

 \it Center for Gravitation and Fundamental
     Metrology, VNIIMS, Ozyornaya St., 46, Moscow 119361, Russia and

 \it Institute of Gravitation and Cosmology, Peoples' Friendship
     University of Russia,  Miklukho-Maklaya St.,6, Moscow 117198,
     Russia

 \end{center}

 \begin{abstract}

 Gravitational $D$-dimensional model with $l$ scalar fields
and several forms is considered. When cosmological type diagonal
metric is chosen, an electromagnetic composite brane ansatz is
adopted and certain restrictions on the branes are imposed the
conformally covariant Wheeler-DeWitt (WDW) equation for the model
is studied. Under certain restrictions asymptotic solutions to WDW
equation are found in the limit of the formation of the billiard
walls which reduce the problem to the so-called quantum billiard
on the $(D+ l -2)$-dimensional Lobachevsky space. Two examples of
quantum billiards are considered. The first one deals with
$9$-dimensional quantum billiard for $D = 11$ model with $330$
four-forms which mimic  space-like $M2$- and $M5$-branes of $D=11$
supergravity. The second one deals with the $9$-dimensional
quantum billiard for  $D =10$ gravitational model with one scalar
field, $210$ four-forms and $120$ three-forms which mimic
space-like $D2$-, $D4$-, $FS1$- and $NS5$-branes  in $D = 10$ $II
A$ supergravity. It is shown that in both examples wave functions
vanish in the limit of the formation of the billiard walls (i.e.
we get a quantum resolution of the singularity for $11D$ model)
but magnetic branes could not be neglected in calculations of
quantum asymptotic solutions  while they are irrelevant for
classical oscillating behaviour  when all $120$ electric branes
are present.

\end{abstract}


 \large

  \section{Introduction}

This paper  deals with  the  quantum billiard approach for
$D$-dimensional \\
cosmological-type  models defined on the
 manifold
 $(u_{-},u_{+}) \times \R^{D-1}$, where $D \geq 4$.

 The billiard approach in classical gravity originally appeared in the
 dissertation of  Chitr\'e \cite{Chit} for the explanation the
 BLK-oscillations \cite{BLK} in the   Bianchi-IX model
 \cite{Mis0,Mis1}. In this approach  a simple triangle billiard
 in the Lobachevsky space $H^2$ was used.

 In \cite{GS}  the billiard approach for $D=4$ was extended to the quantum
 case. Namely, the solutions to  the Wheeler-DeWitt (WDW) equation \cite{DeWitt} were  reduced
to the problem of finding the spectrum of the Laplace-Beltrami
operator on a Chitr\'e's triangle billiard. Such approach  was also
used in \cite{Kir-94} in the context of studying the large scale
inhomogeneities  of the metric in the vicinity of the singularity.

 A straightforward generalization of the Chitr\'e's billiard to the multidimensional case
 was performed in  \cite{IKM1,IKM2,IMb0} where multidimensional cosmological model with
 multicomponent  ``perfect'' fluid  and $n$ Einstein factor spaces was  studied.
 In   \cite{IMb0} the search of oscillating behaviour near the singularity was
 reduced to the problem of proving the finiteness of
 the billiard volume. This  problem  was reformulated in terms of
 the problem of the illumination of the  sphere $S^{n-2}$ by point-like sources.
  In   \cite{IMb0} the inequalities on the Kasner parameters were
 found  and the ``quantum billiard'' approach was considered; see also \cite{IMb0-q,Kir-95}.
 The classical  billiard approach for multidimensional models
 with  fields of forms and scalar fields  was suggested in  \cite{IMb1}, where the
 inequalities for the Kasner parameters were also written. For certain examples these inequalities
 have played a key role in the proof of the never-ending oscillating   behaviour
 near the singularity which takes place in  effective gravitational models
 with forms and scalar fields induced by superstrings  \cite{DamH1,DH1,DHN}.
  It was shown  in \cite{DamH3} that in these  models the  parts of billiards
  are related to Weyl chambers of certain
  hyperbolic Kac-Moody (KM)  Lie algebras \cite{Kac,Sac,BS,HPS}.
  This fact  simplifies  the proof of the finiteness of the billiard
  volume. Using this approach the well-known result
  from \cite{DHSp}   on the critical dimension
  of pure gravity  was explained using hyperbolic algebras in \cite{DamHJN}. For
  reviews on the billiard approach see   \cite{DHN,IMb-09}.

 In recent publications  \cite{KKN,KN} the quantum billiard approach for the multidimensional
 gravitational model with several forms was considered. The main motivation for this approach is coming
 from  the quantum  gravity paradigm;  see \cite{N,K}  and references therein.
 It should be noted that the asymptotic solutions to the WDW equation presented
 in these papers are equivalent to the solutions  obtained earlier in
 \cite{IMb0}. The wave function  ($\Psi_{KKN}$) from  \cite{KKN,KN} corresponds to the harmonic time
 gauge, while the wave function ($\Psi_{IM}$) from  \cite{IMb0} is related to
 the  ``tortoise'' time gauge. (These functions are connected by a certain conformal transformation
 $\Psi_{KKN} = \Omega \Psi_{IM}$.)  In  \cite{IMb0,KKN,KN} a ``semi-quantum'' approach was
 used: the gravity (of a toy model) was quantized but the matter sources (e.g. fluids, forms)
 were considered at the classical level.
\fnm[3]\fnt[3]{Here, one should also mention the recent papers by Lecian (e.g. with co-authors)
  \cite{Lec1,Lec2,Lec3,Lec4} devoted to the quantum billiard approach in the Mixmaster model which
 were inspired by refs.  \cite{KKN,KN}.}
 Such a semi-quantum form of the WDW equation  for the model with fields of forms and
 a scalar field was suggested earlier in \cite{LMMP}.

 In our previous publication \cite{IMqb-1}  we have  used another
 form of the WDW equation with enlarged  minisuperspace which includes the form potentials
 \cite{IMJ}. We have suggested another version of  the quantum billiard approach
 by   deducing the asymptotic solutions to WDW equation for
  the model with fields of forms when a non-composite electric brane ansatz
  has been  adopted.

  In \cite{IMqb-1} we have  considered an example of a $9$-dimensional quantum  billiard
 for $D = 11$ model with $120$ four-forms  which mimic space-like $M2$-brane solutions
 ($SM2$-branes) in  $D=11$ supergravity.\fnm[4]\fnt[4]{For $S$-brane solutions see
  \cite{S1,S2,I-sbr,Ohta} and refs. therein.}
 It was  shown in \cite{IMqb-1} that  the wave  function vanishes  as $y^0 \to -
 \infty$ (i.e. at the singularity),  where $y^0$ is the ``tortoise''
 time-like  coordinate in the minisuperspace.

 In this paper we substantially generalize the approach of \cite{IMqb-1} to the case when
 scalar (dilatonic) fields and dilatonic couplings are added into consideration.
 Here the composite electromagnetic ansatz for branes is considered instead of
 non-composite electric one from \cite{IMqb-1}.  We present new
 examples of  quantum billiards with electric and magnetic $S$-branes in $D=11$
 and $D=10$ models, which are non-composite analogues of truncated
 bosonic sectors of  $D=11$ and $D= 10$ supergravitational models.
 In both examples of billiards  magnetic branes do not participate in
 the formation of the billiard walls since magnetic walls are
 hidden by electric ones. The adding of magnetic branes does
 not change the classical asymptotic oscillating behaviour of scale factors
 and scalar field (for $D=10$).
 In the quantum case  adding of magnetic
 branes changes   the asymptotic behaviour of the wave function,
 but nevertheless, as in \cite{IMqb-1},  the wave  function vanishes  as $y^0 \to -
 \infty$.  For $D = 11$ this means a quantum resolution of the
 singularity for the model with electric and magnetic branes which mimic
 (space-like)  $SM2$- and $SM5$-branes in $11D$ supergravity.

  \section{The setup}

 Here we study  the multidimensional gravitational
 model governed by the  action
 \bear{2.1}
  S_{act} = \frac{1}{2\kappa^{2}}
  \int_{M} d^{D}z \sqrt{|g|} \{ R[g]  - h_{\alpha \beta}
  g^{MN} \partial_{M} \varphi^\alpha \partial_{N} \varphi^\beta
  \\ \nn
   - \sum_{a \in \Delta}
  \frac{\theta_a}{n_a!} \exp[ 2 \lambda_{a} (\varphi) ] (F^a)^2_g
  \}
    + S_{YGH},
 \ear
where $g = g_{MN}(z) dz^{M} \otimes dz^{N}$ is a metric on the
manifold $M$, ${\dim M} = D$, $\varphi=(\varphi^\alpha)\in \R^l$
is a vector from dilatonic scalar fields,
 $(h_{\alpha\beta})$ is a non-degenerate symmetric
 $l\times l$ matrix ($l\in \N$),
 $\theta_a  \neq 0$, and we have
 $$F^a =  dA^a   =\frac{1}{n_a!} F^a_{M_1 \ldots M_{n_a}}
  dz^{M_1} \wedge \ldots \wedge dz^{M_{n_a}}
 $$
which is an $n_a$-form ($n_a \geq 2$) on  $M$ and $\lambda_{a}$ is a
$1$-form on $\R^l$ :
 $\lambda_{a} (\varphi) =\lambda_{a \alpha}\varphi^\alpha$,
 $a \in \Delta$, $\alpha=1,\ldots,l$.
In (\ref{2.1})
we denote $|g| = |\det (g_{MN})|$,
  $(F^a)^2_g =
   F^a_{M_1 \ldots M_{n_a}} F^a_{N_1 \ldots N_{n_a}}
   g^{M_1 N_1} \ldots g^{M_{n_a} N_{n_a}},$
 $a \in \Delta$, where $\Delta$ is some finite set of (colour) indices
 and $S_{\rm YGH}$ is the
standard (York-Gibbons-Hawking) boundary term \cite{Y,GH}. In the
models with one time and the usual fields of forms all $\theta_a > 0$
when the signature of the metric is $(-1,+1, \ldots, +1)$. For
such a choice of signature
 $\theta_b < 0$ corresponds to a ``phantom'' form field $F^b$.

\subsection{Ansatz for composite brane configurations}

We  consider the manifold
 \beq{2.10}
  M = (u_{-}, u_{+})  \times \R^{n},
 \eeq
with the metric
 \beq{2.11}
  g=  w e^{2{\gamma}(u)} du \otimes du   +
  \sum_{i=1}^{n} e^{2\phi^i(u)} \varepsilon(i) dx^i \otimes dx^i ,
 \eeq
where  $w = \pm1$, $\varepsilon(i)  =
\pm 1$,  $i=1,\ldots,n$.  The dimension of $M$ is $D = 1 + n$.
Here  one may replace $\R^{n}$ in (\ref{2.10}) by $\R^{k} \times
(S^1)^{n-k}$, $0 \leq k \leq n$, without any change of all
the relations as presented below.

Although in what follows all examples deal with cosmological
($S$-brane) solutions with $w = - 1$ and $\varepsilon(i)  = + 1$
for all $i$, we reserve here  general notations for signs just
keeping in mind possible future applications to static
configurations with $w = 1$, $\varepsilon(1)  = -1$ and
$\varepsilon(k) = + 1$ for $k
> 1$ (e.g. fluxbranes, wormholes etc.) and solutions with several
time-like directions.

By $\Omega = \Omega(n)$  we denote a set of all non-empty subsets
of $\{ 1, \ldots,n \}$. The number of elements in $\Omega$ is
$|\Omega| = 2^n - 1$.

For any $I = \{ i_1, \ldots, i_k \} \in \Omega$, $i_1 < \ldots <
i_k$, we use the following notations:
 \bear{2.16}
  \tau(I) \equiv dx^{i_1}  \wedge \ldots \wedge dx^{i_k},
  \\
  \label{2.17}
  \eps(I) \equiv \eps(i_1) \ldots \eps(i_k),  \\
   \label{2.19}
   d(I) = |I| \equiv k .
 \ear

For fields of forms we consider the following composite electromagnetic
ansatz:
 \beq{2.1.1}
  F^a= \sum_{I\in\Omega_{a,e}}{\cal F}^{(a,e,I)}+
          \sum_{J\in\Omega_{a,m}}{\cal F}^{(a,m,J)},
 \eeq
where
 \bear{2.1.2}
  {\cal F}^{(a,e,I)}=d\Phi^{(a,e,I)}\wedge\tau(I), \\
  \label{2.1.3}
  {\cal F}^{(a,m,J)}= e^{-2\lambda_a(\varphi)}*(d\Phi^{(a,m,J)}
  \wedge\tau(J))
 \ear
 are elementary forms of electric and magnetic types, respectively,
 $a\in\tri$, $I\in\Omega_{a,e}$, $J\in\Omega_{a,m}$ and
 $\Omega_{a,v} \subset \Omega$, $v = e,m$.

 In (\ref{2.1.3})
 $*=*[g]$ is the Hodge operator on $(M,g)$.

For scalar functions we put
 \beq{2.1.5}
   \varphi^\alpha=\varphi^\alpha(u), \quad
   \Phi^s=\Phi^s(u),
 \eeq
 $s\in S$. Thus, $\varphi^{\alpha}$ and $\Phi^s$ are functions on $(u_{-}, u_{+})$.

Here and below
 \beq{2.1.6}
  S=S_e \sqcup S_m, \quad
  S_v=\sqcup_{a\in\tri}\{a\}\times\{v\}\times\Omega_{a,v},
 \eeq
 $v=e,m$ and  $\sqcup$ is the union
of non-intersecting sets.
The set $S$ consists of elements $s=(a_s,v_s,I_s)$,
where $a_s \in \tri$ is the colour index, $v_s = e, m$ is the electromagnetic
index, and the set $I_s \in \Omega_{a_s,v_s}$ describes the location
of the brane.

Due to (\ref{2.1.2}) and (\ref{2.1.3})
 \beq{2.1.7}
  d(I)=n_a-1, \quad d(J)=D-n_a-1,
 \eeq
for $I\in\Omega_{a,e}$ and $J\in\Omega_{a,m}$, $a\in \tri$, i.e.
in electric and magnetic case, respectively.

 \subsection{Sigma-model action}

Here we present two restrictions on the sets of branes which
guarantee the diagonal form of the  energy-momentum tensor and the
existence of the sigma-model representation (without additional
constraints) \cite{IMC} (see also \cite{IMtop}).

 The first restriction deals with any pair of two (different) branes both
 electric ($ee$-pair) or magnetic ($mm$-pair) with coinciding color index:
   \beq{2.2.2a}
  {\bf (R1)} \quad d(I \cap J) \leq d(I)  - 2,
  \eeq
 for any $I,J \in\Omega_{a,v}$, $a\in \tri$, $v= e,m$ (here $d(I) =
 d(J)$).

  The second restriction deals with any pair of two branes
  with the same color index, which include one electric and one magnetic
  brane ($em$-pair):
 \beq{2.2.3a}
  {\bf (R2)} \quad d(I \cap J) \neq 0,
  \eeq
 where  $I \in \Omega_{a,e}$, $J \in\Omega_{a,m}$, $a\in\tri$.

 These restrictions are satisfied identically
 in the non-composite case, when there are no two branes
 corresponding to the same form $F^a$ for any $a \in \tri$.

It follows from \cite{IMC} that the equations of motion for the model
 (\ref{2.1}) and the Bianchi identities,
   $ d{\cal F}^s=0$,
   $s \in S_m$, for fields from (\ref{2.11}),
  (\ref{2.1.1})--(\ref{2.1.5}), when restrictions ${\bf (R1)}$ and
${\bf (R2)}$ are  imposed, are equivalent to the equations of motion
for the $\sigma$-model governed by the action
 \beq{2.2.7}
  S_{\sigma 0} =  \frac{\mu}{2}
  \int du {\cal N}  \biggl\{\hat G_{AB}
  \dot \sigma^A  \dot \sigma^B
  + \sum_{s\in S}\eps_s \exp{(-2U_A^s\sigma^A)}
  (\dot \Phi^s)^2 \biggr\},
 \eeq
where $\dot x\equiv dx/du$, $(\sigma^A)=(\phi^i,\varphi^\alpha)$,
$\mu \neq 0$, the index set  $S$ is defined in (\ref{2.1.6}) and
${\cal N}=\exp(\gamma_0-\gamma)>0$ is modified lapse function with
$\gamma_0(\phi)  \equiv \sum_{i=1}^n \phi^i$,
 \beq{2.2.9}
  (\hat G_{AB})=\barr{cc}
  G_{ij}& 0\\
  0& h_{\alpha\beta}
 \earr,
 \eeq
is a truncated target space metric with
  \beq{2.2.10}
   G_{ij}=  \delta_{ij} - 1,
  \eeq
and co-vectors
 \beq{2.2.11}
    U_A^s \sigma^A = \sum_{i \in I_s} \phi^i -
  \chi_s \lambda_{a_s}(\varphi),
  \quad
  (U_A^s) =  (\delta_{iI_s}, -\chi_s \lambda_{a_s \alpha}),
 \eeq
 $s=(a_s,v_s,I_s)$.

 Here $\chi_e=+1$ and $\chi_m=-1$;
 \beq{2.2.12}
  \delta_{iI}=\sum_{j\in I}\delta_{ij}
 \eeq
 is the indicator of $i$ belonging
 to $I$: $\delta_{iI}=1$ for $i\in I$ and $\delta_{iI}=0$ otherwise; and
 \beq{2.2.13a}
   \eps_s=\eps(I_s) \theta_{a_s} \ {\rm for} \ v_s = e; \qquad
   \eps_s = -\eps[g] \eps(I_s) \theta_{a_s} \ {\rm for} \ v_s = m,
 \eeq
  $s\in S$, $\eps[g]\equiv\sign\det(g_{MN})$.

 In the electric case $({\cal F}^{(a,m,I)}=0)$
 when any factor space with the coordinate $x^i$ is compactified to
 a circle of length $L_i$,  the action (\ref{2.2.7}) coincides with the action
(\ref{2.1}) if  $\mu=-w/\kappa_0^2$, $\kappa^{2} = \kappa^{2}_0
L_1 \ldots L_n$.

 In what follows we will use the scalar products
 of $U^s$-vectors $(U^s,U^{s'})$; $s,s' \in S$, where
 \beq{3.1.1}
  (U,U')=\hat G^{AB} U_A U'_B,
 \eeq
 for $U = (U_A), U' = (U'_A) \in \R^{N_0}$, $N_0 = n + l$ and
 \beq{3.1.2}
  (\hat G^{AB})=\left(\begin{array}{cc}
  G^{ij}&0\\
  0&h^{\alpha\beta}
  \end{array}\right)
 \eeq
is the matrix inverse to  the matrix
 (\ref{2.2.9}). Here
 \beq{3.1.3}
    G^{ij}= \delta^{ij} +\frac1{2-D},
 \eeq
 $i,j=1,\dots,n$.

The scalar products (\ref{3.1.1})  read \cite{IMC}
 \beq{3.1.4}
  (U^s,U^{s'})=d(I_s\cap I_{s'})+ \frac{d(I_s)d(I_{s'})}{2-D}+
  \chi_s\chi_{s'}\lambda_{a_s \alpha} \lambda_{a_{s'}
  \beta} h^{\alpha \beta},
 \eeq
where $(h^{\alpha\beta})=(h_{\alpha\beta})^{-1}$ and
$s=(a_s,v_s,I_s)$,  $s'=(a_{s'},v_{s'},I_{s'})$ belong to $S$.

The action (\ref{2.2.7}) may also be written in the form
 \beq{4.1.6}
   S_\sigma=\frac{\mu}{2} \int du{\cal N}\left\{
  {\cal G}_{\hat A\hat B}(X)\dot X^{\hat A}\dot X^{\hat B} \right\},
  \eeq
where $X = (X^{\hat A})=(\phi^i,\varphi^\alpha,\Phi^s)\in
 {\R}^{N}$, $N = n +l + m$, $m = |S|$ is the number of branes
  and minisupermetric    ${\cal G}=
   {\cal G}_{\hat A \hat B}(X)dX^{\hat A}\otimes dX^{\hat B}$
   on the minisuperspace  ${\cal M}= \R^{N}$ is defined as
   follows:
 \beq{3.2.3n}
   ({\cal G}_{\hat A\hat B}(X))=\left(\begin{array}{ccc}
   G_{ij}&0&0\\[5pt]
         0&h_{\alpha\beta}&0\\[5pt]
        0&0&\eps_s \exp(-2U^s(\sigma))\delta_{ss'}
 \end{array}\right).
 \eeq

 The minisuperspace metric (\ref{3.2.3n}) may  also be written in
 the form
  \beq{3.2.3g}
  {\cal G}=\hat G+\sum_{s\in S}\eps_s
  \e^{-2U^s(\sigma)}d\Phi^s\otimes d\Phi^s,
  \eeq
where
 \bear{4.1.8}
   \hat G=\hat G_{AB}d \sigma^A
   \otimes d \sigma^B=G_{ij}d\phi^i\otimes d\phi^j+
   h_{\alpha \beta} d\varphi^{\alpha} \otimes d\varphi^{\beta},
 \ear
is truncated minisupermetric and $U^s(\sigma)=U_A^s \sigma^A$ is defined in
 (\ref{2.2.11}).

In what follows we denote
 \beq{4.1.11}
   U^{\Lambda}(\sigma)=U_A^{\Lambda} \sigma^A= \gamma_0(\phi),
   \qquad (U_A^{\Lambda})=(U^{\Lambda}_i = 1, U^{\Lambda}_{\alpha} = 0).
 \eeq
This vector is time-like and all $(U^{s},U^{\Lambda}) < 0$, since
\cite{IMC}
 \beq{4.1.15}
    (U^{\Lambda},U^{\Lambda})=-\frac{D-1}{D-2},
    \qquad (U^{s},U^{\Lambda})= \frac{d(I_s)}{2-D} .
 \eeq

\section{Quantum billiard approach}

In this section  we develop a quantum analogue of the billiard
approach which deals with asymptotical solutions to Wheeler-DeWitt
(WDW) equation.

\subsection{Restrictions.}

First we outline  restrictions on parameters which will be used in
derivation of the ``quantum billiard''
  \bear{3.1.4u}
  (i) \quad (U^s,U^{s})  > 0, \\ \label{3.1.4h}
  (ii) \quad \quad (h_{\alpha \beta}) > 0, \\ \label{3.1.4e}
  (iii) \quad \quad \quad \eps_s > 0,
  \ear
 for all $s$.  
 These restrictions  are necessary conditions for the formation
 of infinite ``wall'' potential in certain limit (see below).
 The first restriction reads (see (\ref{3.1.4})) 
 \beq{3.1.uu}
   (U^s,U^{s})=    d(I_s) \left(1 +\frac{d(I_{s})}{2-D} \right) +
  \lambda_{a_s \alpha} \lambda_{a_{s}  \beta} h^{\alpha \beta} >0.
 \eeq
The second restriction means that the matrix
 $(h_{\alpha \beta})$ is positive definite, i.e.
 the so-called phantom scalar fields are not considered.

\subsection{Wheeler-DeWitt  equation}

 Now we  fix the temporal gauge as follows:
 \beq{4.2.1}
  \gamma_0-\gamma= 2f(X),  \quad  {\cal N} = e^{2f},
 \eeq
where $f$: ${\cal M}\to \R$ is a smooth function. Then we obtain
the Lagrange system with the Lagrangian
 \beq{4.2.3}
   L_f = \frac{\mu}{2} \e^{2f}{\cal G}_{\hat A\hat B}(X)
   \dot X^{\hat A}\dot X^{\hat B}
 \eeq
and the energy constraint
 \beq{4.2.4}
  E_f = \frac{\mu}{2} \e^{2f}{\cal G}_{\hat A\hat B}(X)
  \dot X^{\hat A}\dot X^{\hat B}=0.
 \eeq

The set of Lagrange equations with the constraint (\ref{4.2.4})
is equivalent to the set of Hamiltonian equations for the
Hamiltonian
 \beq{4.2.3h}
   H^f=\frac{1}{2 \mu} \e^{-2f}{\cal G}^{\hat A\hat B}(X)
    P_{\hat A}  P_{\hat B}
 \eeq
 with the constraint
 \beq{4.2.4h}
   H^f= 0,
 \eeq
where $ P_{\hat A} = \mu \e^{2f}{\cal G}_{\hat A\hat B}(X)  \dot
X^{\hat B}$ are momenta (for fixed gauge) and $({\cal G}^{\hat
A\hat B}) = ({\cal G}_{\hat A\hat B})^{-1}$.

Here we use the  prescriptions of covariant and
conformally covariant quantization of the hamiltonian constraint
$H^f= 0$ which was suggested initially by Misner \cite{Mis} and
considered afterwards in \cite{Hal,IMZ,HK1,HK2} and some other
papers.

We obtain the Wheeler-DeWitt (WDW) equation\fnm[5]\fnt[5]{For the WDW
equation in $4D$ gravity see \cite{DeWitt}.}
 \beq{4.2.5}
   \hat{H}^f \Psi^f \equiv
    \left(-\frac{1}{2\mu}\Delta\left[e^{2f}{\cal G}\right]+
    \frac{a}{\mu} R\left[e^{2f}{\cal G}\right]
     \right)\Psi^f=0,
  \eeq
where
 \beq{4.2.5a}
  a=a_N= \frac{(N-2)}{8(N-1)},
 \eeq
  $N = n+l + m$.

Here $\Psi^f = \Psi^f(X)$ is the wave function corresponding to
the $f$-gauge (\ref{4.2.1}) and satisfying the relation
\fnm[6]\fnt[6]{We eliminate here a typo in a corresponding
 relation from \cite{IMJ}.}
 \beq{4.2.7}
  \Psi^f= e^{bf} \Psi^{f=0}, \quad b = b_N =(2-N)/2.
 \eeq

 In (\ref{4.2.5}) we denote by $\Delta[{\cal G}^f]$ and
 $R[{\cal G}^f]$  the Laplace-Beltrami operator and the scalar
 curvature corresponding to the metric
 \beq{4.2.7G}
 {\cal G}^f =   e^{2f} {\cal G},
 \eeq
 respectively.

 The choice of minisuperspace covariant form for the Hamiltonian operator $\hat{H}^f$
 (\ref{4.2.5})  with  arbitrary real number $a$  is one of the
 solutions to the operator ordering problem in multidimensional quantum cosmology
 \cite{ChrZ2,ChrZ2a,HP,ChrZ3}. The Lapalace-Beltrami form of WDW
 equation was considered previously in \cite{DeWitt,Kuh,HPT,ChrZ1,ChrZ1a}.
 Similar prescription  appears   in quantization of a point-like
 particle moving in a curved background, for a review  see \cite{Tag1,Tag2}.

 It was shown in \cite{HK1,HK2} by rigorous constraint quantization
 of parametrized relativistic gauge systems  in curved spacetimes that
 the  privileged choice for $a$ in  cosmological case  is given by (\ref{4.2.5a}).
  For this  value of $a$ and $N > 1$ there is one-to-one correspondence between solutions
 to WDW equations for any two choices of  temporal gauges given by
 (\ref{4.2.1}) with  smooth functions $f_1$ and $f_2$ instead of $f$, respectively.
 This fact follows from (\ref{4.2.7}) and the following relation:
 \beq{4.2.8}
 \hat{H}^f =   e^{-2f} e^{bf} \hat{H}^{f=0} e^{-bf}.
 \eeq
We note that the coefficients $a_N$ and $b_N$ are the well-known ones
in the conformally covariant theory of a scalar field \cite{BD}.

 Now we put $f = f(\sigma)$. Then we get
 \bear{4.2.9}
   \tri[{\cal G}^f]
    =  \e^{\bar{U}}|\bar{G}|^{-1/2} \frac\partial{\partial \sigma^A}\left(\bar{G}^{AB}
     \e^{- \bar{U}} |\bar{G}|^{1/2} \frac\partial{\partial \sigma^B}\right)
      \\ \nonumber
       +  \sum_{s\in S} \e^{2 \bar{U}^s(\sigma)}
        \left(\frac\partial{\partial\Phi^s}\right)^2,
  \ear
 where
   \beq{4.2.9U}
   \bar{U}= \sum_{s\in S} \bar{U}^{s}, \qquad  \bar{U}^{s} = U^{s}(\sigma) - f
   \eeq
    and
     \beq{4.2.9G}
     \bar{G}_{AB} =  e^{2f} \hat{G}_{AB}, \qquad \bar{G}^{AB} =  e^{-2f}
     \hat{G}^{AB},
     \eeq
     $|\bar{G}| = |\det{(\bar{G}_{AB})}|$.

   Here we deal with a special class of asymptotical solutions to
   WDW-equation. Due to restrictions (\ref{3.1.4h}) and (\ref{3.1.4e})
 the  (minisuperspace) metrics  $\hat{G}$, $\cal{G}$ have a pseudo-Euclidean
   signatures $(-,+,...,+)$. We put
       \beq{5.1}
         e^{2f} = - (\hat{G}_{AB}\sigma^A \sigma^B)^{-1},
       \eeq
   where we impose  $\hat{G}_{AB}\sigma^A \sigma^B < 0$.

   Here  we  use a diagonalization of  $\sigma$-variables
       \beq{5.1.z}
      \sigma^{A} = S^{A}_{a}z^{a},
       \eeq
   $a = 0, ..., N_0-1$,  with $N_0 = n +l$, obeying $\hat{G}_{AB} \sigma^{A} \sigma^{B}
    = \eta_{ab} z^{a}z^{b}$, where $(\eta_{ab}) =
    {\diag}(-1,+1,...,+1)$.

    We restrict  the WDW equation to the lower light cone
    $V_{-} = \{z = (z^{0}, \vec{z}) | z^{0} < 0, \eta_{ab} z^{a}z^{b} < 0 \}$
   and  introduce the Misner-Chitr\'e-like coordinates
    \bear{5.2z}
     z^{0} = - e^{-y^{0}}\frac{1 + \vec{y}^{2}}{1 -\vec{y}^{2}}, \\
     \label{5.2zz}
     \vec{z} = - 2 e^{-y^{0}} \frac{\vec{y}}{1 - \vec{y}^{2}},
     \ear
    where $y^{0} < 0$ and $\vec{y}^{2} < 1$.
    In these variables we have $f = y^{0}$.

  Using the relation $f_{,A} =  \bar{G}_{AB} \sigma^{B}$,
 following from   (\ref{5.1.z}), we obtain
  \beq{5.3}
       \tri[\bar{G}]f = 0, \qquad \bar{G}^{AB} f_{,A} f_{,B} = -1.
  \eeq
 These relations just follow from
 the  relation
   \beq{5.4}
      \bar{G} = - dy^{0} \otimes d y^{0} + h_L,
   \eeq
   where
   \beq{5.5}
       h_L =  \frac{4 \delta_{rs} dy^{r} \otimes dy^{s}}{(1 - \vec{y}^{2})^2},
  \eeq
  (the summation over $r,s = 1,..., N_0 -1$ is assumed). Here the
  metric $h_L $ is defined on the unit ball
  $D^{N_0 -1} = \{ \vec{y} \in \R^{N_0 -1}| \vec{y}^{2} < 1
  \}$.  $D^{N_0 -1}$  with the metric $h_L$ is a realization of
  the $(N_0 -1)$-dimensional  Lobachevsky space $H^{N_0 -1}$.

 For the wave function  we suggest the following ansatz:
      \beq{5.6}
         \Psi^f = e^{C(\sigma)} \Psi_{*},
       \eeq
   where  the prefactor $e^{C(\sigma)}$ is chosen for the sake of
   cancellation the terms linear in derivatives $\Psi_{*,A}$ arising in
   calculation of   $\tri[{\cal G}^f] \Psi^f$. This takes place
   for
      \beq{5.7}
       C = C(\sigma) = \frac{1}{2} \bar{U} =
       \frac{1}{2}(\sum_{s\in S} U^{s}_A \sigma^A - m f).
      \eeq
  With this choice of the prefactor we obtain
     \bear{5.8}
     \e^{\bar{U}}|\bar{G}|^{-1/2} \frac\partial{\partial \sigma^A}\left(\bar{G}^{AB}
     \e^{- \bar{U}} |\bar{G}|^{1/2} \frac\partial{\partial
     \sigma^B}\right) (e^{C(\sigma)} \Psi_{*}) \\ \nn
     = e^{C(\sigma)} [\tri[\bar{G}] \Psi_{*} +
      (\tri[\bar{G}] C) \Psi_{*} - \bar{G}^{AB} c_{,A} c_{,B} \Psi_{*}].
      \ear

   Using relations  (\ref{5.3}) we get
    \bear{5.9}
    \tri[\bar{G}] C = \frac{1}{2} (n+l-2) \sum_{s \in S}
    U^s_A \sigma^A,
      \\ \label{5.10}
      \bar{G}^{AB} c_{,A} c_{,B} =
      \frac{1}{4} [ e^{-2f} \sum_{s, s' \in S} (U^s,U^{s'})
        \\ \nn
       - 2m  \sum_{s \in S} U^s_A
      \sigma^A   - m^2 ].
     \ear

  The calculation of the scalar curvature  $R\left[e^{2f}{\cal
  G}\right]$  gives us  the following formula:
   \bear{5.11}
     R \left[e^{2f}{\cal G}\right] =
          e^{-2f}
     [ - \sum_{s \in S} (U^s,U^{s})  - \sum_{s, s' \in
     S} (U^s,U^{s'})]
        \\ \nn
     + 2 (N - 1) \sum_{s \in S} U^s_A \sigma^A
       + (N-1)(m + 2 - n -l).
      \ear

  Collecting relations (\ref{5.9}), (\ref{5.10}) and (\ref{5.11})
  we obtain  the following identity:
     \bear{5.12}
    \left(-\frac{1}{2}\Delta\left[e^{2f}{\cal G}\right]+
    a  R\left[e^{2f}{\cal G}\right]
     \right) (e^{C(\sigma)} \Psi_{*}) =
  \\ \nn
     e^{C(\sigma)}
     \left(-\frac{1}{2}\Delta\left[\bar{G} \right] -
      \frac{1}{2} \sum_{s\in S} \e^{2 \bar{U}^s}
       \left(\frac\partial{\partial\Phi^s}\right)^2 + \delta V \right)
       \Psi_{*},
  \ear
  where
     \beq{5.13}
      \delta V = A e^{-2f} - \frac{1}{8} (n+l-2)^2.
      \eeq

  Here we denote
          \beq{5.14}
      A  =   \frac{1}{8(N-1)} [ \sum_{s, s' \in
     S} (U^s,U^{s'}) - (N - 2) \sum_{s \in S} (U^s,U^{s}) ].
      \eeq
  In what follows we call $A$ as $A$-number.

  It should be noted that linear in $\sigma^{A}$ terms, which appear in
  (\ref{5.9}), (\ref{5.10}) and (\ref{5.11}) are canceled in (\ref{5.12}) due to
  our choice of conformal coupling $a =(N-2)/(8(N-1))$.

   Now we put
      \beq{5.15}
        \Psi^f = e^{C(\sigma)} e^{iQ_s \Phi^s} \Psi_{0,L}(\sigma),
      \eeq
  where the parameters $Q_s \neq 0$  correspond  to the charge densities
  of branes and $e^{iQ_s \Phi^s} = \exp(i \sum_{s \in S} Q_s \Phi^s)$.
  Using  (\ref{5.12}) we get

  \bear{5.16}
  \hat{H}^f \Psi^f = \mu^{-1} e^{C(\sigma)}
  e^{iQ_s \Phi^s}  \left(-\frac{1}{2} \tri[\bar{G}]+ \qquad \qquad  \right.\\ \nn
    \left.  \frac{1}{2} \sum_{s \in S} Q_s^2 e^{-2f + 2U^s(\sigma)}
    + \delta V \right) \Psi_{0,L}=0.
  \ear
 Here and in what follows   $U^s(\sigma) = U^s_A \sigma^A$.

\subsection{Asymptotic behavior of solutions for $y^0 \to - \infty$}

Now  we proceed with the studying the  asymptotical solutions to
WDW equation in the limit $y^0 \to - \infty$. Due to (\ref{5.15})
and (\ref{5.16}) this equation reads
  \beq{5.17}
    \left(-\frac{1}{2} \tri[\bar{G}]+
    \frac{1}{2} \sum_{s \in S} Q_s^2 e^{-2f + 2U^s(\sigma)}
    + \delta V  \right) \Psi_{0,L}=0.
  \eeq

It was shown in  \cite{IMb1} that
  \beq{5.18}
    \frac{1}{2} \sum_{s \in S} Q_s^2 e^{-2f + 2U^s(\sigma)}
    \to V_{\infty},
  \eeq
as $y^0 = f  \to - \infty$.
   Here  $V_{\infty}$ is the potential of infinite walls which
   are produced by branes
      \beq{5.18a}
   V_{\infty} = \sum_{s \in S} \theta_{\infty}( \vec{v}_s^2 -1 - (\vec{y} - \vec{v}_s)^2),
      \eeq
  where we denote $\theta_{\infty}(x) = + \infty $, for $x \geq
  0$ and $\theta_{\infty}(x) = 0$ for $x < 0$. The vectors
  $\vec{v}_s$, $s \in S$, which belong to $\R^{N_0 -1}$ ($N_0 = n
  +l$),   are defined by  
      \beq{5.20}
      \vec{v}_s =  -  \vec{u}_s/u_{s0},
      \eeq
  where the $N_0$-dimensional vectors  $u_s = (u_{s0},\vec{u}_s) = (u_{sa})$
  are obtained from $U^s$-vectors using the diagonalization matrix $(S^{A}_{a})$
  from  (\ref{5.1.z})
  \beq{5.21}
    u_{sa} = S^{A}_{a} U^s_A.
  \eeq
    Due to (\ref{3.1.4u}) we get
  \beq{5.21a}
  (U^s,U^s) = -(u_{s0})^2 + (\vec{u}_s)^2 > 0
  \eeq
  for all $s$. In what follows
  we use a diagonalization (\ref{5.1.z}) from
  \cite{IMb1} obeying
  \beq{5.21b}
   u_{s0} > 0
  \eeq
  for all $s \in S$. The inverse matrix   $(S_{A}^{a}) =
  (S^{A}_{a})^{-1}$ defines
   the map which is inverse to (\ref{5.1.z})
   \beq{5.21c}
   z^{a} = S_{A}^{a} \sigma^{A},
   \eeq
  $a = 0, ..., N_0 -1$.
     The inequalities (\ref{5.21a})
  imply  $|\vec{v}_s| > 1$ for all $s$. The potential $V_{\infty}$
  corresponds to the billiard $B$ in the multidimensional Lobachevsky
  space $(D^{N_0 -1}, h_L)$. This  billiard is an open domain in
 $D^{N_0 -1}$ obeying the a set of inequalities:
     \beq{5.22}
       |\vec{y} - \vec{v}_s| < \sqrt{\vec{v}_s^2 -1} = r_s,
     \eeq
  $s \in S$. The boundary of the billiard  $\partial B$ is formed by  parts of
  hyper-spheres with  centers in $\vec{v}_s$ and radii $r_s$.

   The condition (\ref{5.21b}) is  obeyed for the
   diagonalization (\ref{5.21c}) with
     \beq{5.21cc}
     z^{0} = U_A \sigma^{A}/\sqrt{|(U,U)|},
     \eeq
   where $U$ is a time-like vector
      \beq{5.21U}
      (U,U) < 0,
     \eeq
    and
    \beq{5.21Us}
      (U,U^s) < 0,
     \eeq
     for all $s \in S$. (Here the  relation $(U,U^s) = -  u_{s0} \sqrt{|(U,U)|}$  is used.)
     The inequalities (\ref{5.21U}) and (\ref{5.21Us})
     are satisfied identically if
        \beq{5.21UL}
         U = k U^{\Lambda}, \qquad k > 0,
           \eeq
     see  (\ref{4.1.15}). This choice
     of $U$ with $k = 1$ was done in \cite{IMb1}.

     {\bf Remark 1.} {\em  Conditions   (\ref{5.21b}) (or (\ref{5.21Us})) may be relaxed.
     In this case we obtain a more general definition of the billiard walls (e.g. for $u_{s0} < 0$  and
     $u_{s0} = 0$) described in \cite{IMb-09}.}

   Thus, we are led to the asymptotical relation for the function
   $\Psi_{0,L}(y^0,\vec{y})$
   \beq{5.23}
    \left(-\frac{1}{2}\tri[\bar{G}]+ \delta V \right) \Psi_{0,L}=0
    \eeq
    with the zero boundary condition $\Psi_{0,L|\p B} = 0$
    imposed.

     Due to (\ref{5.4}) we get
     $\tri[\bar{G}] = - (\p_0)^2 +
     \tri[h_L]$, where $\tri[h_L] = \Delta_{L}$ is
     the Lapalace-Beltrami operator corresponding to the
     Lobachevsky metric $h_L$.

     By separating the variables,
      \beq{5.24}
    \Psi_{0,L}= \Psi_{0}(y^0) \Psi_{L}(\vec{y}),
    \eeq
    we obtain  the following asymptotical relation (for $y^{0} \rightarrow -
    \infty$)
     \beq{5.25}
     \left( \left(\frac{\partial}{\partial y^{0}}\right)^{2}  +
     2Ae^{-2y^{0}} +  E - \frac{1}{4} (N_0 -2)^2 \right)\Psi_{0} =
     0,
     \eeq
      where
       \beq{5.26}
       \Delta_{L}\Psi_{L} = - E \Psi_{L}, \qquad  \Psi_{L|\p B}=0.
       \eeq

       We assume that the minus Laplace-Beltrami
       operator $(-\Delta_{L})$  with the zero
       boundary conditions has a spectrum obeying
       the following inequality:
       \beq{5.27}
        E \geq  \frac{1}{4} (N_0 -2)^2.
        \eeq
         The examples of  billiards obeying this restriction
         were considered in \cite{KKN,KN} (see also the next section).

        Here we  restrict ourselves to the case of negative
        $A$-number
        \beq{5.27a}
         A < 0.
         \eeq

        Solving equation (\ref{5.25}) we get for $A < 0$
        the following set of basis solutions:
       \beq{5.28}
       \Psi_{0} = {\cal B}_{i \omega} \left(\sqrt{2|A|}e^{-y^{0}}\right),
        \eeq
      where  ${\cal B}_{i \omega}(z) = I_{i \omega}(z), K_{i\omega}(z)$ are the modified Bessel
      functions and
      \beq{5.29}
       \omega = \sqrt{E -  \frac{1}{4} (N_0 -2)^2} \geq 0.
       \eeq

       By using the asymptotical relations
        \beq{5.30}
         I_{\nu} \sim \frac{e^z}{\sqrt{2 \pi z}}, \qquad  K_{\nu} \sim \frac{e^{-z}}{\sqrt{2 z}},
         \eeq
        for $z \to + \infty$, we find
         \beq{5.31}
         \Psi_{0} \sim C_{\pm} \exp \left( \pm \sqrt{2|A|}e^{-y^{0}} + \frac{1}{2} y^{0} \right)
         \eeq
         for $y^0 \to - \infty$. Here  $C_{\pm}$ are non-zero
         constants, ``plus'' corresponds to ${\cal B} = I$ and ``minus'' -
         to ${\cal B} = K$.

         Now we evaluate the prefactor $e^{C(\sigma)}$ in
             (\ref{5.15}), where
             \beq{5.32}
             C(\sigma) = \frac{1}{2}(U(\sigma) - m f).
             \eeq
          Here we denote
            \beq{5.33}
            U(\sigma) = U_A \sigma^A =  \sum_{s\in S} U^{s}_A
            \sigma^A, \qquad U_A  =  \sum_{s\in S} U^{s}_A.
            \eeq

          In what follows we use the vector $U = (U_A )$ as a time-like vector in the relation for  $z^0$  in
          (\ref{5.21cc}).  Thus, we need to impose the restriction (\ref{5.21U}) ($(U,U) < 0$) .

          Using (\ref{5.2z}), (\ref{5.21cc})  and $f = y^0$
          we obtain
            \beq{5.34}
             C(\sigma) = \frac{1}{2}(q z^0  - m f) =
              \frac{1}{2} \left( - q e^{-y^{0}}
              \frac{(1 + \vec{y}^{2})}{1 -\vec{y}^{2}}  - m y^0
              \right),
             \eeq
            where
             \beq{5.34q}
             q = \sqrt{-(U,U)} > 0.
             \eeq

            Combining relations (\ref{5.15}), (\ref{5.24}), (\ref{5.31}) and (\ref{5.34})
            we get
             \beq{5.35}
             \Psi^f
              \sim C_{\pm} \exp \left( \theta^{\pm}(|\vec{y}|)e^{-y^{0}} - \frac{1}{2}(m -1) y^{0}
              \right) e^{iQ_s \Phi^s} \Psi_{L}(\vec{y}),
              \eeq
             as $y^0 \to - \infty$  for any fixed $\vec{y} \in B$ and  $C_{\pm} \neq 0$.
             Here we denote
              \beq{5.36}
              \theta^{\pm}(|\vec{y}|) = - \frac{q}{2}   \frac{(1 + \vec{y}^{2})}{(1 -\vec{y}^{2})}
                                          \pm \sqrt{-2A},
              \eeq
              where ``plus'' corresponds to the solution with ${\cal B} = I$ and ``minus'' -
              to ${\cal B} = K$.

         Relation (\ref{5.14}) may be rewritten in the following
         form:
         \beq{5.14A}
         A  =   \frac{1}{8(N-1)} [ (U,U) - (N - 2) \sum_{s \in S} (U^s,U^{s}) ].
             \eeq
        where we have used the identity
        \beq{5.37}
        (U,U) = \sum_{s, s' \in  S} (U^s,U^{s'})
         \eeq
      following from the definition of $U$ in   (\ref{5.33}). It should be noted that
       restrictions $(U,U) < 0$ and $(U^s,U^{s}) > 0$, $s \in S$, imply $A <
                 0$.

      Now we study the asymptotical behaviour of the wave function  (\ref{5.15})

       \beq{5.37Psi}
        \Psi^f = e^{C(\sigma)} e^{iQ_s \Phi^s} {\cal B}_{i \omega}
          \left(\sqrt{2|A|}e^{-y^{0}}\right) \Psi_{L}(\vec{y}),
           \eeq
       with  $C(\sigma)$ from (\ref{5.34}) and $(U,U) <
            0$, $A < 0$.

       {\bf A.} Let  ${\cal B} = K$. Then
         \beq{5.37P}
        \Psi^f \to 0
         \eeq
       as $y^0 \to - \infty$  for fixed $\vec{y} \in B$ and $\Phi^s \in \R$, $s \in
          S$.

      This  follows just from the asymptotic relation (\ref{5.35}).

       {\bf B.} Now we consider the case  ${\cal    B} = I$.

          {\bf B1.} First we put
           \beq{5.38A}
           \quad  \frac{1}{2} q > \sqrt{2|A|},
           \eeq
         or, equivalently,
        \beq{5.38U}
         \sum_{s \in S} (U^s,U^{s}) < -(U,U).
           \eeq
         We get  $$\Psi^f \to 0 $$
        as $y^0 \to - \infty$  for fixed $\vec{y} \in B$
        and $\Phi^s \in \R$, $s \in  S$.      This also follows from (\ref{5.35}).
        The equivalence of the     conditions  (\ref{5.38A}) and  (\ref{5.38U})
        could be readily verified using the relations
       (\ref{5.34q}) and (\ref{5.14A}).

        {\bf B2.} Let
       \beq{5.38Aa}
         \quad \frac{1}{2} q = \sqrt{2|A|},
        \eeq
       or, equivalently,
         \beq{5.38Uu}
        \sum_{s \in S} (U^s,U^{s}) = -(U,U).
           \eeq
        Then we also get $$\Psi^f \to 0 $$
         as $y^0 \to - \infty$  for fixed $\vec{y}
            \in B \setminus \{ \vec{0} \} $ and $\Phi^s \in \R$, $s \in
         S$.  This  also follows from   (\ref{5.35}).

       {\bf B3.} Now we consider the third case
       \beq{5.38Ab}
      \frac{1}{2} q < \sqrt{2|A|},
       \eeq
      or, equivalently,
        \beq{5.38Ub}
        \sum_{s \in S} (U^s,U^{s}) > -(U,U).
      \eeq
        Let the point $ \vec{0}$ belong to the billiard
        $B$ (this is valid when relations (\ref{5.21b}) are satisfied)
        and $\Psi_{L}(\vec{0}) \neq 0$.
        Then there exists $\eps >0$ such that for all $\vec{y}$
        obeying $|\vec{y}| < \eps$
        and all $\Phi^s \in \R$, $s \in  S$)
       \beq{5.39}
        |\Psi^f| \to + \infty
       \eeq
          as $y^0 \to - \infty$.

      Indeed,  since the billiard $B$ is an open domain, $\vec{0} \in B$
   and $\Psi_{L}(\vec{y})$ is continuous  function
   there exists $\eps > 0$ such that the ball $B_{\eps} =  \{ \vec{y}| |\vec{y}| < \eps \}$
   belongs to $B$ and $\Psi_{L}(\vec{y}) \neq 0$ for all $ \vec{y} \in B_{\eps}$.
    Due to (\ref{5.38Ub}) $\eps > 0$  may be chosen such
   that
          \beq{5.38Ay}
      \frac{1}{2} q \frac{(1 + \vec{y}^{2})}{1 -\vec{y}^{2}}  < \sqrt{2|A|},
       \eeq

  for all $|\vec{y}| < \eps$. Then relations (\ref{5.35}), (\ref{5.38Ay})
  and  $\Psi_{L}(\vec{y}) \neq 0$    imply (\ref{5.39}) for all
  $ \vec{y} \in B_{\eps}$.

 {\bf Remark 2.}
  {\em It should be noted that solution (\ref{5.28})  is
  similar to those which were found in quantum cosmological
  models with $\Lambda$-term, perfect fluid etc.,
  see \cite{BIMZ,IM-95} and references therein. For these solutions
  we have $v =e^{q z^0}$ instead of $e^{-y^0}$, where $v$
  is the volume \cite{BIMZ} or the ``quasi-volume'' \cite{IM-95} scale factor.
  Our  restriction $A < 0$ corresponds to the restriction $\Lambda < 0$ for
  the solutions from \cite{BIMZ}. }

\section{Examples}

Here we illustrate our approach by two examples of quantum
billiards in dimensions $D = 11$ and  $D = 10$. In what follows we
use the notation $\Omega(n,k)$ for the set of all subsets of  $\{ 1,
\ldots,n \}$, which contain $k$ elements .
 Any element of $\Omega(n,k)$ has the form
 $I = \{ i_1, \ldots, i_k \}$, $1 \leq i_1 < \ldots < i_k \leq n$,
The number of elements in $\Omega(n,k)$ is $C_n^k = \frac{n!}{k!
(n-k)!}$.

In this section we deal with $(n+ 1)$-dimensional cosmological
metric of Bianchi-I type,

\beq{6.0}
  g=  - e^{2{\gamma}(u)} du \otimes du   +
  \sum_{i=1}^{n} e^{2\phi^i(u)} dx^i \otimes dx^i,
 \eeq
where $ u \in (u_{-}, u_{+})$.

\subsection{$9$-dimensional billiard in $D=11$ model}

Let us consider an $11$-dimensional gravitational model with several
4-forms, which produce non-composite analogues of $SM$-brane
solutions in $D = 11$ supergravity \cite{CJS}. The action  reads
as follows:
 \beq{6.1}
  S_{11} = \frac{1}{2\kappa^2_{11}}
  \int_{M} d^{11}z \sqrt{|g|} \{ {R}[g] + {\cal L}  \}+ S_{YGH},
 \eeq

where first we put ${\cal L}  =  {\cal L}_e$, where
  \beq{6.2}
  {\cal L}_e =
   -  \frac{1}{4!} \sum_{I \in \Omega (10,3)}  (F^I_{4,e})^2_g.
   \eeq

Here $F^I_{4,e}$ is ``electric'' 4-form with the index $I \in
\Omega(10,3)$. The number of such forms is $C_{10}^3 = 120$.

The action (\ref{6.1}) with  $\cal{L}$ from (\ref{6.2}) describes
non-composite analogues of  $SM2$-brane solutions which are given
by the  metric  (\ref{6.0}) with $n=10$ and
 \beq{6.2e}
 F^I_{4,e} =
  {\cal F}^{(a,e,I)},
  \eeq
$I \in \Omega(10,3)$, $a = (4,e,I)$, where electric monoms ${\cal
F}^{(a,e,I)}$ are defined in (\ref{2.1.2}).

 Consider the
 non-trivial case when all charge densities of branes $Q_s$,
 $s \in S_e$, are non-zero. In the classical case we get
 a $9$-dimensional billiard $B \in H^9$ with $120$ ``electric''
walls \cite{IMqb-1}. This classical billiard  coincides with
the $9d$ billiard from \cite{DamH3,DHN}. $B$ has a finite volume.
It is a union of several identical ``small'' billiards which have
finite volumes as they correspond to the Weyl chamber of the
hyperbolic Kac-Moody algebra $E_{10}$ \cite{DamH3}.

{\bf Remark 3.} {\em  In \cite{IMqb-1} we have used $120$ form
fields and non-composite ansatz for branes to avoid the appearance
of the set of $45$ constraints which arise for composite solutions
with diagonal metric \cite{IMC,IMS}. These constraints are coming
from the relations $T_{ij} = 0$, $1 \leq i < j \leq 10$, where
$T_{ij}$ are spatial components of the stress-energy tensor. We
note that in \cite{DHN,DamH3} this problem was circumvented by
considering non-diagonal metrics from the very beginning.}

Let us calculate $(U,U)$, where $U = U_e = \sum_{s \in S_e}
 U^s$. We get $U_i = \sum_{I \in \Omega(10,3)} \delta_{iI}$, where
 $\delta_{iI}$  was defined in (\ref{2.2.12}). Thus, $U_i$ is the
number of sets
 $I \in \Omega(10,3)$ which contain  $i$ ($i = 1, \dots, 10$). It is obvious
 that $U_i = C_9^2 = 36$. Here $U = 36 U^{\Lambda}$ (see (\ref{4.1.11}))
 and hence we may use  the $z$-variables from \cite{IMb0,IMb1}.

 Then we get (see (\ref{3.1.3}))
  \beq{6.3}
   (U,U) = G^{ij}U_i U_j =  \sum_{i,j = 1}^{10} (\delta^{i j} -
   \frac{1}{9})(36)^2= - 1440 < 0
   \eeq
in agreement with our restriction (\ref{5.21U}). Since $N = 130$
($m=120$) and $(U^s,U^s) =2$ we obtain from  (\ref{5.14A}) the
following value for the $A$-number \cite{IMqb-1}:
        \beq{6.3A}
         A = A_e(M2) = - \frac{1340}{43}.
        \eeq

 In this case the inequality  (\ref{5.38U}) is satisfied
 identically since \cite{IMqb-1}
   \beq{6.4}
    240 = \sum_{s \in S_e} (U^s,U^{s}) < -(U,U) = 1440.
   \eeq

 The minus Laplace-Beltrami
 operator $(-\Delta_{L})$ on $B$ with the zero
 boundary conditions imposed has a spectrum obeying
  restriction (\ref{5.27}) with $N_0 = 10$ \cite{KN}.

 We get from the previous analysis the asymptotical vanishing of
 the wave function $\Psi^f \to 0$ as $y^0 \to - \infty $.

Now we consider the electromagnetic case, which mimics  solutions with
$SM2$- and $SM5$-branes.

We put in  (\ref{6.1}) ${\cal L} =  {\cal L}_e + {\cal   L}_m$,
where
 \beq{6.5}
  {\cal L}_m =
   -  \frac{1}{4!} \sum_{J \in \Omega (10,6)}  (F^J_{4,m})^2_g.
   \eeq

  Here $F^J_{4,m}$ is a ``magnetic'' 4-form with the index $J \in \Omega(10,6)$.
The number of such forms is $C_{10}^6 = 210$. We extend the
cosmological electric ansatz by adding  the following relations:
\beq{6.2m}
 F^J_{4,m} = {\cal F}^{(a,m,J)},
  \eeq
$J \in \Omega(10,6)$, $a = (4,m,J)$, where  ${\cal F}^{(a,m,J)}$
are defined in (\ref{2.1.3}). For charge densities we put $Q_s
\neq 0$, $s \in S$.

In the electromagnetic case we get the same  $9$-dimensional
billiard $B \in H^9$ as in the electric case, since
 magnetic walls are hidden by electric ones. This could be readily
 verified using the billiard chamber
  $$W =
   \{ \sigma | \hat{G}_{AB}\sigma^A \sigma^B < 0;
   U(\sigma) < 0; U^s(\sigma) < 0, s \in S  \}$$
 belonging to the lower light  cone
 and  the fact that any magnetic $U$-vector is the sum of
 two electric ones. The  Lobachevsky space $H^{9}$ may be
 identified with the hypersurface $y^0 =0$ in the lower light
 cone. Then the billiard  $B$ may be obtained
 just by  the projection of  $W$ onto $H^{9}$: $(y^0,\vec{y}) \mapsto  \vec{y}$.
 Adding into our consideration a magnetic brane with $U_m = U_{1e} + U_{2e}$, where
  $U_{1e}$ and $U_{2e}$ correspond to electric  branes,
 gives  a new inequality in the definition of $W$: $U_m(\sigma) < 0$,
 which is satisfied identically due to relations $U_{1e}(\sigma) < 0$ and
 $U_{2e}(\sigma) < 0$ from the  definition of  $W$ in the electric case.
 Thus, the addition  of any magnetic $SM5$-brane does not change
 the electric billiard chamber  nor  the electric billiard.

A calculation similar to the electric case  of $U = U_{em} =
\sum_{s \in S}  U^s$, $s \in S$, gives us $U_i = 36 + 126 = 162$,
where now $126 = C_9^5$ is the number of sets
 $J \in \Omega(10,6)$ which contain  $i$ ($i = 1, \dots, 10$).

  We obtain
  \beq{6.6}
   (U,U) = G^{ij}U_i U_j =  -  \frac{10}{9} (162)^2= - 29160 < 0.
   \eeq

Now we have  $m = 330$, $N = 340$ and $(U^s,U^s) =2$ for all $s$.
We get from (\ref{5.14A}) the following value for the $A$-number
        \beq{6.3em}
         A = A_{em}(M2,M5) = - \frac{12940}{113}.
        \eeq

 In this case the inequality  (\ref{5.38U}) is also obeyed
   \beq{6.4em}
    660 = \sum_{s \in S} (U^s,U^{s}) < -(U,U) = 29160.
   \eeq

 The analysis carried out in the previous section
 implies the asymptotical vanishing of
 the wave function $\Psi^f \to 0$ as $y^0 \to - \infty $.

Thus, adding  $210$ magnetic $SM5$-branes does not change the
``electric'' billiard $B$ and the  spectrum of the
Laplace-Beltrami  operator $\Delta_{L}$ on $B$ (with zero boundary
condition). At quantum level we get a quantitatively different
behaviour $\Psi^f \to 0$ as $y^0 \to - \infty $, since parameters
of the solutions $q$ and $\sqrt{2|A|}$ in electric and
electromagnetic cases are different: $q_{em}/q_e = 9/2$ and
$\sqrt{|A_{em}|}/\sqrt{|A_{e}|} \sim 1.9$. Hidden magnetic walls
change the asymptotical behaviour of $\Psi^f $ (see (\ref{5.35}))
though at the classical level they could be neglected.

{\bf Remark 4.} {\em  The wave function,
 corresponding to the harmonic gauge also vanishes, i.e.
 $\Psi = e^{-by^0} \Psi^f \to 0$ as $y^0 \to - \infty $,
 since the term $(-by^0)$ in the
 exponent is suppressed  by the $e^{-y^{0}}$-term from
   (\ref{5.35}).}

\subsection{$9$-dimensional billiard in $D=10$ model}

Now we consider a $10$-dimensional gravitational model with one
scalar field and several 4- and 3-forms. This model  gives us
non-composite analogues of space-like $D2$-, $ $FS1$-, D4$- and
$NS5$-brane solutions in $D = 10$ $II A$ supergravity.

The action  reads as follows:
 \beq{7.1}
  S_{10} = \frac{1}{2\kappa^2_{10}}
  \int_{M} d^{11}z \sqrt{|g|} \{ {R}[g] -
     g^{MN} \partial_{M} \varphi^\alpha \partial_{N} \varphi^\beta
   + {\cal L}  \}+ S_{YGH}.
 \eeq

 First we put ${\cal L}  =  {\cal L}_e$, where
  \beq{7.2}
  {\cal L}_e =
   -  \frac{1}{4!} e^{ 2 \lambda_{4} \varphi }
     \sum_{I_1 \in \Omega (9,3)}  (F^{I_1}_{4,e})^2_g
   -  \frac{1}{3!} e^{2 \lambda_{3} \varphi}
     \sum_{I_2 \in \Omega (9,2)}  (F^{I_2}_{3,e})^2_g.
   \eeq

Here $F^{I_1}_{4,e}$ is the ``electric'' $4$-form, $I_1 \in
\Omega(9,3)$, and $F^{I_2}_{3,e}$ is the ``electric''  $3$-form, $I_2
\in \Omega(9,2)$. The number of
 $4$-forms is $C_9^3 = 84$ and the number of
 $3$-forms is $C_9^2 = 36$. In  (\ref{7.2})  $\lambda_{4} =
 \frac{1}{2 \sqrt{2}}$ and $\lambda_{3} = -2 \lambda_{4}$.

The action (\ref{7.1}) with  $\cal{L}$ from (\ref{7.2}) describes
non-composite $SD2$-, $SFS1$-brane solutions which are given by
the  metric  (\ref{6.0}) with $n=9$ and
 \beq{7.2e}
  F^{I_1}_{4,e} =
  {\cal F}^{(a_1,e,I_1)},  \qquad F^{I_2}_{3,e} =
  {\cal F}^{(a_2,e,I_2)},
  \eeq
 with $I_1 \in \Omega(9,3)$, $a_1 =(4,e,I_1)$ and $I_2 \in \Omega(9,2)$,
  $a_2 =(3,e,I_2)$, see (\ref{2.1.2}). We put $Q_s \neq 0$, $s \in S_e$.

 In  the classical case we get the same  $9$-dimensional billiard
 $B \in H^9$ with $120$ ``electric'' walls
 as in the $SM2$-brane case \cite{DH1,DHN}.

Let us us calculate $(U,U)$, where $U = U_e = \sum_{s \in S_e}
 U^s$. We get

\beq{7.3e}
 U_i = \sum_{I_1 \in \Omega(9,3) } \delta_{iI_1} + \sum_{I_2 \in
 \Omega(9,2)}  \delta_{i I_2}=   C_8^2 + C_8^1 = 36.
\eeq

 The first term in the sum $C_8^2 = 28$ is the number
 of sets $I_1 \in \Omega(9,3)$ which contain  $i$  and
 the second term $C_8^1 = 8$ is the number
 of sets $I_2 \in \Omega(9,2)$ which contain  $i$ ($i = 1, \dots, 9$).
 Thus, $U_i = 36$ for all $i$. For the $\varphi$-component we get
 (see (\ref{2.2.11}))

 \beq{7.3le}
  U_{\varphi} = - 84 \lambda_4 - 36 \lambda_3 =-12 \lambda_4.
 \eeq

 Then we get  the same value
  \beq{7.3}
   (U,U) = G^{ij}U_i U_j + U_{\varphi}^2 =  \sum_{i,j = 1}^{9} (\delta^{i j} -
   \frac{1}{8})(36)^2 + \frac{(12)^2}{8} = - 1440 < 0
   \eeq
as in the $SM2$-case.

 Since $N = 130$ ($m=120$) and $(U^s,U^s) =2$, $s \in S_e$,
 we obtain from  (\ref{5.14A}) the
 same value for the $A$-number as in the
 $SM2$-case
        \beq{7.4}
         A = A_e(D2,FS1) = - \frac{1340}{43}
        \eeq
and the asymptotical vanishing of
 the wave function $\Psi^f \to 0$ as $y^0 \to - \infty $.

Now we consider the electromagnetic case, which mimics  solutions
with
 $SD2$-, $SFS1$-, $SD4$-  and $SNS5$-branes  in $D = 10$ $II
A$ supergravity.

We put in  (\ref{7.1}) ${\cal L} =  {\cal L}_e + {\cal   L}_m$,
where
 \beq{7.5}
  {\cal L}_m =
   -  \frac{1}{4!} e^{ 2 \lambda_{4} \varphi } \sum_{J_1 \in \Omega (9,5)}  (F^{J_1}_{4,m})^2_g
   -  \frac{1}{3!} e^{ 2 \lambda_{3} \varphi } \sum_{J_2 \in \Omega (9,6)}  (F^{J_2}_{3,m})^2_g.
   \eeq

  Here $F^{J_1}_{4,m}$,  $J_1 \in \Omega(9,5)$, is the ``magnetic''
   4-form and $F^{J_2}_{3,m}$,  $J_2 \in \Omega(9,6)$, is the ``magnetic''
   3-form.

The number of ``magnetic''  4-forms is $C_{9}^5 = 126$, while the
number of ``magnetic''  3-forms is $C_{9}^6 = 84$. We extend the
cosmological electric ansatz by adding  the following relations:
 \beq{7.6}
 F^{J_1}_{4,m} = {\cal F}^{(a_1,m,J_1)},
 \qquad F^{J_2}_{3,m} = {\cal F}^{(a_2,m,J_2)},
 \eeq
$J_1 \in \Omega(9,5)$, $a_1 = (4,m,J_1)$ and  $J_2 \in
\Omega(9,6)$, $a_2 = (3,m,J_2)$, where ${\cal F}^{(a,m,J)}$ are
defined in  (\ref{2.1.3}). For the charge densities we put $Q_s \neq
0$, $s \in S$.

 In the electromagnetic case  the $9$-dimensional billiard
 is the same as in the pure electric case, i.e. $B =
  B_e \in H^9$, since  magnetic walls are hidden by  electric ones.
 (This may be readily proved along a similar line to that was followed
 for $M$-branes in the previous subsection.)

The calculation of $(U,U)$ in the electromagnetic case
 $U = U_{em} = \sum_{s \in S}  U^s$ gives

\beq{7.3em}
 U_i = 36 + C_8^4 + C_8^5 = 162.
\eeq

 Here $C_8^4 = 70$ is the number
 of sets $I_1 \in \Omega(9,5)$ which contain  $i$ and
  $C_8^5 = 56$ is the number
 of sets $I_2 \in \Omega(9,6)$ which contain  $i$ ($i = 1, \dots, 9$).
 Thus, $U_i = 162$ for all $i$. For the $\varphi$-component we get
 (see (\ref{2.2.11}))

 \beq{7.3lem}
  U_{\varphi} = - 12 \lambda_4 +  126 \lambda_4 + 84 \lambda_3
   = -54 \lambda_4.
 \eeq

  We obtain
  \bear{7.6U}
   (U,U) = G^{ij}U_i U_j +  U_{\varphi}^2 =  -  \frac{9}{8}(162)^2
   + \frac{1}{8} (54)^2 = - 29160 < 0.
   \ear

Since  $m = 330$, $N = 340$ and $(U^s,U^s) =2$ for all $s$, we get
from (\ref{5.14A}) the following value for the $A$-number
        \beq{7.7}
         A = A_{em}(D2,FS1,D4,NS5) = - \frac{12940}{113}.
         \eeq

Thus, we are led to  the same values of the scalar product $(U,U)$
and the $A$-number as for the model which mimics $SM2$- and
$SM5$-branes. For the wave function we obtain the asymptotic
vanishing $\Psi^f \to 0$ as $y^0 \to - \infty $.
\fnm[7]\fnt[7]{Recently, analogous behaviour of the wave function
was obtained in   \cite{DSp} for $D =4$ simple supergravity, when
a tachyon case was considered.}

{\bf Remark 5.} {\em The coincidence of the $A$-numbers
 $$A_{em}(D2,FS1,D4,NS5) = A_{em}(M2,M5)$$
 is not surprising since
 there is  a one-to-one correspondence between the sets of space-like branes:
 $(SM2,SM5)$ and $(SD2,SFS1,SD4,SNS5)$, which preserves the scalar
 products $(U^s,U^{s'})$. The number $N$ is the same in both
 cases.}

\section{Conclusions}

Here we have continued our approach from \cite{IMqb-1} by
considering the quantum billiard for the cosmological-type model
with $n$ 1-dimensional factor spaces in the theory with several
forms and $l$ scalar fields. After adopting the electromagnetic
composite brane ansatz with certain restrictions on brane
intersections and parameters of the model we have deduced the
Wheeler-DeWitt (WDW) equation for the model, written in the
conformally covariant form. It should be noted that in our
previous paper \cite{IMqb-1} we were dealing with a gravitational
model which contains  fields of forms without scalar fields. In
\cite{IMqb-1} only electric non-composite configurations of branes
were considered. Thus, the generalization of the model from
\cite{IMqb-1} is rather evident.

By imposing  certain restrictions on the parameters of the model we
have obtained the asymptotic solutions to the WDW equation, which are
of the quantum billiard  form since they are governed by the
spectrum of the Lapalace-Beltrami operator on the billiard with
the zero boundary condition imposed. The billiard is a part of the
$(N_0-1)$-dimensional Lobachevsky space $H^{N_0 -1}$, where $N_0 =
n + l$.

Here we have presented two examples of quantum billiards: (a) the
quantum $9d$ billiard for $11D$ gravitational model with $120$
``electric'' $4$-forms and $210$ ``magnetic'' $4$-forms
 which mimics the quantum billiard with space-like $M2$-  and $M5$-branes in $D=11$
 supergravity,
 (b)  the quantum $9d$ billiard for $10D$ gravitational model with one scalar
 field, $84$
``electric'' $4$-forms,  $126$ ``magnetic'' $4$-forms, $36$
``electric'' $3$-forms and  $84$ ``magnetic'' $3$-forms,
 which mimics the quantum billiard with space-like $D2$-, $D4$-, $FS1$- and
 $NS5$-branes in $D=10$ $IIA$ supergravity.

 In both cases we have shown  the asymptotic vanishing of the basis wave functions  $\Psi^f  \to 0$,
 as $y^0 \to - \infty $,   for any choice of the Bessel function ${\cal B} = K, I$.
 For $D=11$ model this result may be interpreted as a quantum resolution of
 the singularity.  It should be noted that in the approach of  \cite{KKN,KN} 
 asymptotic (basis) solutions to WDW equation in the harmonic gauge are vanishing as $\rho = e^{-y^0} \to +
 \infty$.

 In the examples presented above the magnetic walls change the asymptotical behaviour of
 the wave function $\Psi^f $. Thus, hidden magnetic walls which  do not contribute to
 the asymptotical behaviour of the classical solutions for $y^0 \to - \infty $
 should be taken into account in the quantum case. This is the first
 lesson from this paper. The second one is related to the use of
 the conformally covariant version of the WDW equation. Here we were
 able to develop the quantum billiard approach for the model with branes only for a special
 conformal choice of the parameter $a =  (N-2)/(8(N-1))$ in the WDW equation, where $N = n+l +m$ and
 $m$ is the number of branes. The study of the asymptotical behaviour of
 the wave function (as $y^0 \to - \infty$) for the non-conformal choice of the parameter  
  $a \neq (N-2)/(8(N-1))$  should be a subject of a separate publication.

 It should be noted that in the two examples presented here we have considered  non-composite
 branes  while initially we had formulated the quantum billiard approach
 for  the composite branes  with rather severe restrictions on brane
 intersections. Unfortunately, these restrictions exclude the possibility of efficiently applying
 the formalism to cosmological  models with diagonal metrics in $11D$ and $10D$ $IIA$
 supergravities (the relaxing of these restrictions will lead to quadratic
 constraints  on the brane charge densities $Q_s$ \cite{IMS}). In the classical case
 this obstacle was avoided in \cite{DHN} by considering
the ADM type approach for  non-diagonal cosmological metrics and
using the Iwasawa decomposition. In this case the Chern-Simons
terms were irrelevant for the classical formation of the billiard
walls \cite{DHN}. But in the quantum case the consideration of the
Chern-Simons contributions needs a separate investigation. This
(and some other topics) may be a subject of future publications.

 \end{document}